\documentclass[sigconf]{acmart}

\usepackage{booktabs} 
\usepackage{enumitem}

\acmYear{2018}
\acmConference[WWW 2018]{The 2018 Web Conference}{April 23--27, 2018}{Lyon, France}
\acmBooktitle{WWW 2018: The 2018 Web Conference, April 23--27, 2018, Lyon, France}

\editor{Jennifer B. Sartor}

\begin{document}
\title{Crowd-Machine Collaboration for Item Screening}

\author{Evgeny Krivosheev \\Bahareh Harandizadeh}
\affiliation{%
  \institution{University of Trento, Italy}
}
\email{first.last@unitn.it}

\author{Fabio Casati}
\affiliation{%
  \institution{University of Trento, Italy and \\Tomsk Polytechnic University, Russia}
}
\email{first.last@unitn.it}

\author{Boualem Benatallah}
\affiliation{%
  \institution{UNSW, Sydney, Australia}
}
\email{boualem@cse.unsw.edu.au}

\begin{abstract}
In this paper we describe how crowd and machine classifier can be efficiently combined to screen items that satisfy a set of predicates. 
We show that this is a recurring problem in many domains, present machine-human (hybrid) algorithms that screen items efficiently and estimate the gain over human-only or machine-only screening in terms of performance and cost.     
\end{abstract}

%
%

\keywords{crowdsourcing, machine learning, hybrid systems, classification}

\maketitle

\section{Background and Motivation}

A frequently occurring classification problem consists in identifying items that pass a set of screening tests (filters).
This is not only common in medical diagnosis but in many other fields as well, from database querying - where we filter tuples based on predicates \cite{optimalWidom2014}, to hotel search - where we filter places based on features of interest \cite{Lan_dynamicfilter_hcomp17}, to systematic literature reviews (SLR) - where we screen candidate papers based on a set of criteria to assess whether they are in scope for the review~\cite{Wallace2017crowdMLshort}.
The goal of this paper is to understand how, given a set of trained classifiers whose accuracy may or may not be known for the problem at hand (for a specific query predicate, hotel feature, or paper topic), we can combine machine learning (ML) and human (H) classifiers to screen items efficiently in terms of cost of querying the crowd, while ensuring an accuracy that is acceptable for the given problem.
To make the paper easier to read and the problem concrete, we take the example of SLR mentioned above, which is rather challenging in that each SLR is different and each filtering predicate (called \textit{exclusion criterion} in that context) could be unique to each SLR (e.g., "exclude papers that do not study adults 85+ years old").
Abundant prior art discusses crowd-based filtering (e.g., \cite{optimalWidom2014,Wallace2017crowdMLshort,Mortensen2016crowd,Krivosheev_hcomp17,Krivosheev_www18,Lan_dynamicfilter_hcomp17}), while research on hybrid classification is still in its infancy. Recent papers address the problem of combining machine and crowd intelligence in crowd-powered bots\cite{savenkov_crqa_hcomp17} as well as in  crowdsourced classification\cite{Wallace2017crowdMLshort}.
The problem we address here differs from prior art in that i) we use the information provided by each kind of classifier (machine and human) to improve the effectiveness of the other kind so that they can be stronger together and ii) we consider a probabilistic model that works on a per filter and per item basis to minimize the overall number of crowd votes. 



\section{Problem Statement and Model} 
We assume in input a set of items $i \in I$ to classify (in our example, these are papers to screen), a set of filters $f \in F$ (paper exclusion criteria), a set of ML or H classifiers $c \in C$, and a loss function $L=k*FE + FI$, modeled as a linear combination of false exclusions FE and false inclusions FI, which may carry a different relative weight $k$ (e.g., most authors consider excluding relevant papers a more costly error than including an irrelevant one).
Each classifier $c = \{ cost, a(f), \rho(f,C) \}  \in C$ is associated with the cost of asking one vote on an (item, filter) pair, with a filter-specific estimated accuracy (a 2x2 confusion matrix capturing probability of correct decisions for positive and negative labels), and with its correlation $\rho$ with other classifiers. 
We specify filter-specific accuracy as we have seen that accuracy can vary greatly based on the filter (exclusion criteria) to be evaluated, for both machines (as studied in the experiments described later) and humans (as we reported in~\cite{Krivosheev_www18}). 
We do not discuss here how to obtain ML classifiers as this is the subject of ample literature: we merely assume they are given, and that we may or may not have information on their accuracy and correlation when applied to specific problem (our set of candidate papers and exclusion criteria).
Consequently, we model accuracy as a beta distribution, where we incorporate prior knowledge if available, else we assign an initial uniform $Beta(1,1)$ distribution for both positively and negatively labeled items. 
To simplify the presentation we assume to have three kinds of classifiers: machines (with zero cost per vote and $Beta(1,1)$ accuracy), crowd (with cost 1 and also $Beta(1,1)$ accuracy), and experts, with expert cost $ec$ to which for simplicity we assign perfect accuracy. Consistently with crowdsourcing literature, we also assume that crowd and experts' opinions are independent, while in general we cannot make this assumption for ML classifiers. 
Our goal is, given quality parameters such as the loss function, to identify a strategy that can efficiently (in terms of cost) query the classifiers available and aggregate results while achieving the quality goals.

\section{Strategies and Experiments}
We base the hybrid machine-crowd classification strategy on modifying the \textit{shortest run} (SR) algorithm, developed for crowd-only classification~\cite{Krivosheev_www18}.  
SR proceeds by obtaining a test dataset $T$ from "expensive" experts (usually 10-20 items) used to filter out low accuracy crowd workers, and by performing a  \textit{baseline run} with the - cheaper - crowd classifying a set of $B$ items (usually 50-100) to estimate crowd accuracy and filter selectivity. 
Based on this estimate, and on the (initially empty) set of votes for the items to be classified, it then decides which filter to apply first to which item, to maximize the probability of screening the item out with few votes. It also estimates the expected cost for crowd classification, leaving items to experts when convenient \cite{Krivosheev_www18}. 
We extend SR because i) it was designed for multi-filter screening and has shown to perform better than baseline algorithms for crowd classification, ii) it has a per paper and per item probabilistic model that can leverage prior knowledge on items and filters, and ML classifiers can provide such knowledge, and iii) the algorithm can work with different sizes for test $T$ and baselines $B$. This is important as test items can help us filter out ML classifiers with an expected accuracy lower than a threshold $\overline{a}$ (to be tuned as discussed later), and the more extensive set of crowd-classified items from the baseline can be used to a) assess independence among ML classifiers (which is necessary if we want to pool votes from ML classifiers with simple algorithms such as majority voting), and b) build an ensemble model out of the ML classifiers where the output of each ML classifier is a feature \cite{Dzeroski2004Stacking}.  
Therefore, the classification strategy proceeds as follows: 

\begin{enumerate}[leftmargin=*,topsep=5pt,parsep=0pt,partopsep=2pt]
\item Obtain gold dataset T from expert and use it to screen ML classifiers and to use as tests questions for crowd workers. 
If we start from a Beta(1,1) uniform prior for a given filter, we know the accuracy distribution goes to a $Beta(1+correct\_answers, 1+failed\_answers)$ which has a known pdf and mean.
\item Perform a baseline run on $B$ items (on all filters), both to estimate crowd accuracy on each filters and to get data for the next step.
\item Compute correlation among ML classifiers and remove classifiers with correlation higher than a threshold $c$ that we tune empirically, so that we can meaningfully use weighted majority voting to combine the opinions of different ML classifiers. 
\item Compute the probability that a filter applies to an item by combining the vote of the ensemble ML classifier (which now has a known accuracy). Treat this value as a prior probability for the (filter, item) pair and continue with SR until items are classified by the crowd or until SR decides they are to be left to experts.
\end{enumerate}

Many variations are possible over this basic scheme, including changing the size of test data $T$ and baseline $B$ 
as well as building a model (e.g., logistic regression) to combine classifiers votes as opposed to using majority voting, leveraging the baseline run. 

\begin{figure}[tbh]
    \centering
    \includegraphics[width=0.42\textwidth]{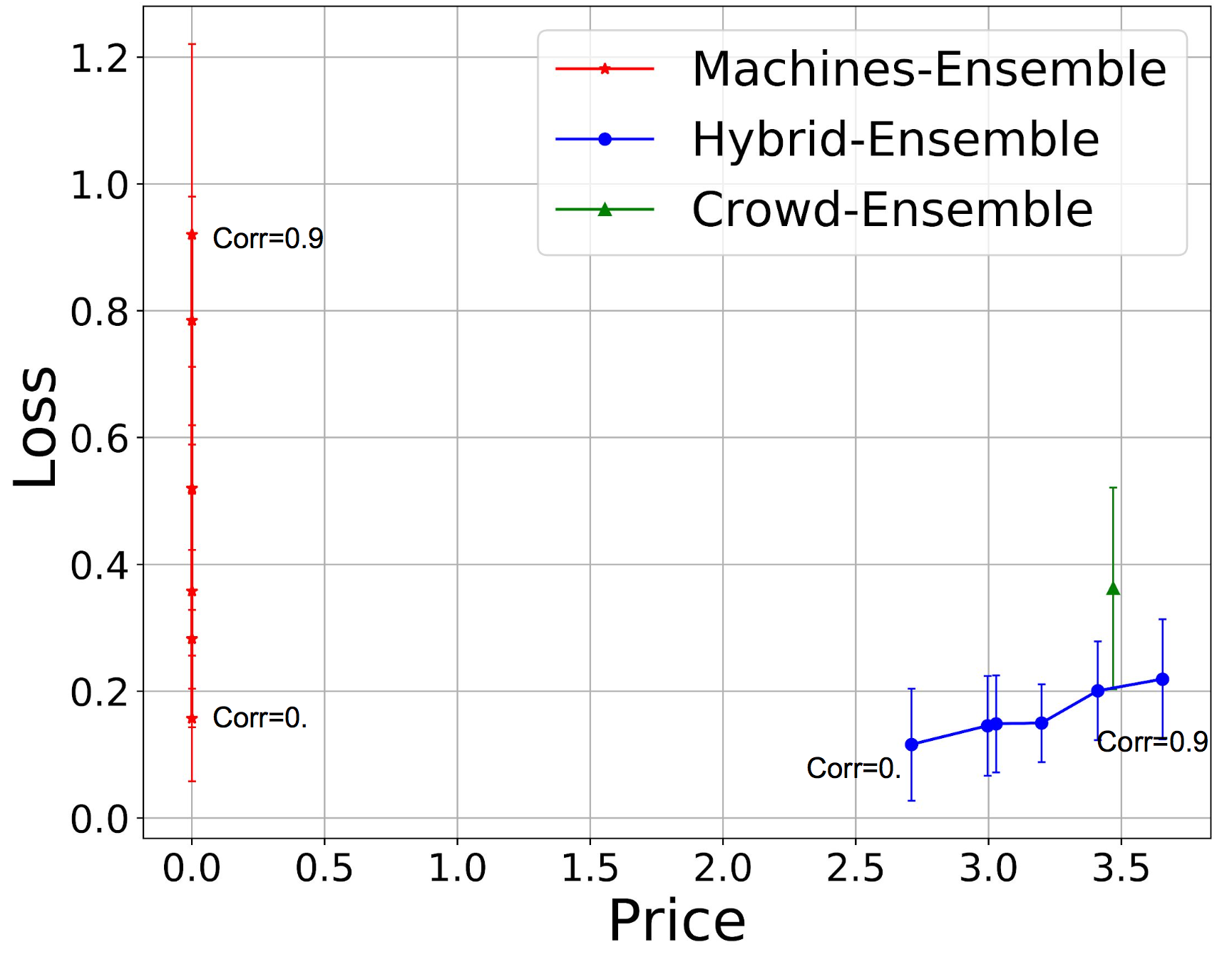}
    \caption{Expected loss vs price for different algorithms. Correlation values between machines = [0., 0.2, 0.3, 0.5, 0.7, 0.9]}
    \label{fig:loss-price}
\end{figure}
\textbf{Experiments.} To assess the approach, we ran experiments on Mechanical Turk (described in \cite{Krivosheev_www18}) as well as leverage existing crowd datasets \cite{Mortensen2016crowd}.
In both cases the experiments are related to SLRs with multiple filters and include over 20000 crowd votes on over 4000  papers. We refer to the cited papers for details on experiment design.
We used these datasets to get realistic data on crowd worker accuracies, on variation of such accuracies by filter and on filter power.
czWe then built classifiers for each filter using a variety of techniques (from KNN to random forest, variations of naive Bayes, and others\footnote{see jointresearch.net for details}) and different sizes of training data to get realistic information on  classifier accuracies and correlations. 
We obtained classifier accuracies in the 0.5-0.95 range and correlations in the 0.2-0.9 range, and crowd accuracy in the 0.55-0.8 range.
We then used this data to simulate a variety of scenarios. 
In Figure~\ref{fig:loss-price} we compare the results of applying
machine only, crowd (with SR), and hybrid strategy, to simulations of classifications for 1000 papers and 4 filters (averages over 50 iterations). 
We simulate 10 ML classifiers with accuracy randomly selected from a 0.5-0.95 range and on which the algorithm assumes no prior knowledge. We screen them with T=20 tests and keep the ones with 0.95 probability of having an accuracy greater than 0.5.  
The threshold for false exclusion error is set at 0.01 as in \cite{Krivosheev_www18} and the weight K in the loss function is set to 5. 
We then plot the average loss and price paid per item as the correlation among classifiers vary from 0 to 0.9 (with price growing with the correlation). 
As we can see, for a similar loss level, hybrid algorithms significantly outperforms the crowd in terms of price, with  savings from 7.4\% to 53.7\% depending on correlation. 
Notice that we disregard here the cost of obtaining the classifier, which, if they  are built from scratch for this specific SLR, needs to be factored in when estimating price and  assessing the best strategy. If classifiers accuracies worsen (e.g., lie in the 0.4-0.8 or 0.3-0.7 range), the savings also decrease approximately by a factor of 2 and 4 respectively. 
For near-zero correlation (very hard to achieve in practice) the ML-only strategy becomes appealing, and then deteriorates. 
The reader can see on the GitHub repository in footnote how  results vary as we change parameters and thresholds.

\textbf{Acknowledgements.} This project has received funding from the EU Horizon 2020 research and innovation programme under the Marie Skodowska-Curie
grant agreement No 690962.

\bibliographystyle{ACM-Reference-Format}
\bibliography{cs} 
\end{document}